# Gaussian Process Regression-Based Lithium-Ion Battery End-of-Life Prediction Model under Various Operating Conditions

Seyeong Park[1], Jaewook Lee[1], Seongmin Heo[*]

*Department of Chemical and Biomolecular Engineering, Korea Advanced Institute of Science and Technology, 291 Daehak-ro, Yuseong-gu, Daejeon 34141, Republic of Korea*

[1]These authors contributed equally.

[*]Corresponding author. E-mail address: smheo@kaist.ac.kr (S. Heo).

## Abstract

For the efficient and safe use of lithium-ion batteries, diagnosing their current state and predicting future states are crucial. Although there exist many models for the prediction of battery cycle life, they typically have very complex input structures, making it very difficult and expensive to develop such models. As an alternative, in this work, a model that predicts the nominal end-of-life using only operating conditions as input is proposed. Specifically, a total of 100 battery degradation data were generated using a pseudo two-dimensional model with three major operating conditions: charging C-rate, ambient temperature and depth-of-discharge. Then, a Gaussian process regression-based model was developed to predict the nominal end-of-life using these operating conditions as the inputs. To improve the model accuracy, novel kernels were proposed, which are tailored to each operating condition. The proposed kernels reduced the lifetime prediction error by 46.62% compared to the conventional kernels.

## Keywords

Lithium-ion battery, End-of-life, Cycle life prediction, Gaussian process regression, Kernel construction

## Copyright

# 1. Introduction

The demand for lithium-ion batteries (LIBs) as energy storage devices is rapidly increasing due to their advantages, such as large capacity and high energy density [1]. LIBs are widely used in products such as electric vehicles, various electronic devices, and large-scale energy storage systems [2]. One of the main challenges for LIBs is ensuring efficient and safe operation, which is often compromised by degradation caused by side reactions such as solid-electrolyte interphase (SEI) formation and lithium plating [3]. As a result, many researchers are focused on accurately diagnosing the battery's current state and predicting its future performance.

The most commonly used metric to represent a battery's current health is the state-of-health (SOH), defined as the ratio between the battery's capacity in its current cycle and its initial capacity [4]. The cycle in which the cell reaches 80% SOH is considered the cell's end-of-life (EOL). Since battery reliability decreases rapidly near the EOL, significant research has focused on accurately predicting this point. Major challenges in EOL prediction arise from the complexity and nonlinearity of battery degradation caused by a complex interplay among different mechanisms [5]. This becomes even more difficult in the case of early prediction, where only data from a few initial cycles are available.

Capacity fade due to degradation mechanisms is affected by two key factors: operating conditions (OCs) and intrinsic cell characteristics. When cells are manufactured using the same process and specifications, the operating conditions largely determine the overall degradation rate, while the final EOL is influenced by cell-to-cell variability, which typically follows a normal distribution with relatively small variances [6]. Among various OCs, the main factors affecting EOL are charging current, state of charge (SOC), and ambient temperature [7, 8].

Methods for battery health prognostics can be broadly categorized into model-based and data-driven methods [9, 10]. Model-based approaches leverage detailed knowledge of electrochemical phenomena within the battery, offering high accuracy at the expense of significant computational resources. On the other hand, data-driven methods aim to identify intrinsic patterns in the data without relying on prior knowledge, making them relatively easy to develop. Such methods are especially favored in battery EOL prediction because they require minimal understanding of complex degradation mechanisms [11, 12].

A key factor in developing data-driven prediction models is the selection of the proper data for making predictions, as the model's accuracy and efficiency depend heavily on the data used. When choosing appropriate data, it is essential to consider which variables are highly correlated with the target variable, and how costly it will be to obtain the data for such variables. Most data-driven studies for EOL prediction rely on operational variables such as voltage, current, temperature, and capacity, collected from cycling tests, to predict battery health [13, 14]. However, obtaining such data can be time-consuming and expensive, making these models inefficient for battery life prediction, as they fail to separate the effects of operating conditions from intrinsic cell characteristics. Therefore, it is crucial not only to determine the right data efficiently but also to build models using a small dataset from as few cycles as possible.

Choosing the proper model is another important factor. Severson et al. [15] effectively predicted the cycle life of LIBs using the elastic net, by appropriately leveraging input features extracted from operational variables such as voltage, current, and temperature. Despite its high predictive performance, many studies prefer deep learning methods, which better capture the nonlinearity and complexity of battery degradation. For example, Ma et al. [16] proposed a model based on long short-term memory to calculate the remaining useful life using time-series data of capacity, demonstrating reduced error in nonlinear datasets. Lee et al. [17] introduced an improved model that accounts for inter- and intra-cycle effects using a 2-dimensional convolutional neural network. While these studies achieved high accuracy, the machine learning models used required a large number of datasets, which presents a significant limitation for real-world battery system applications.

Gaussian process regression (GPR) can be an alternative, as it is known to achieve high performance with a small amount of data at low computational cost [9, 18, 19]. Also, unlike general machine learning techniques, the predicted value can be obtained in the form of a distribution, which has the great advantage of quantifying uncertainty. While GPR has been utilized in several studies to predict battery capacity and EOL, it has yet to account for the effects of different operating conditions effectively [20-22].

To this end, this study proposes a GPR-based model for EOL prediction, which can effectively capture the effects of operating conditions. To avoid using cycling data, the EOL attributable solely to operating conditions is defined as nominal EOL. Only the operating conditions are used as the inputs in the proposed model for its prediction. In particular, the effects of C-rate, ambient temperature, and depth-of-discharge (DOD), which are highly influential on EOL, are utilized for EOL prediction. The proposed model can be considered as a physics-informed model, as it utilizes some well-known relationships between operating conditions and EOL as the prior knowledge. The overall structure of this study is illustrated in Fig. 1.

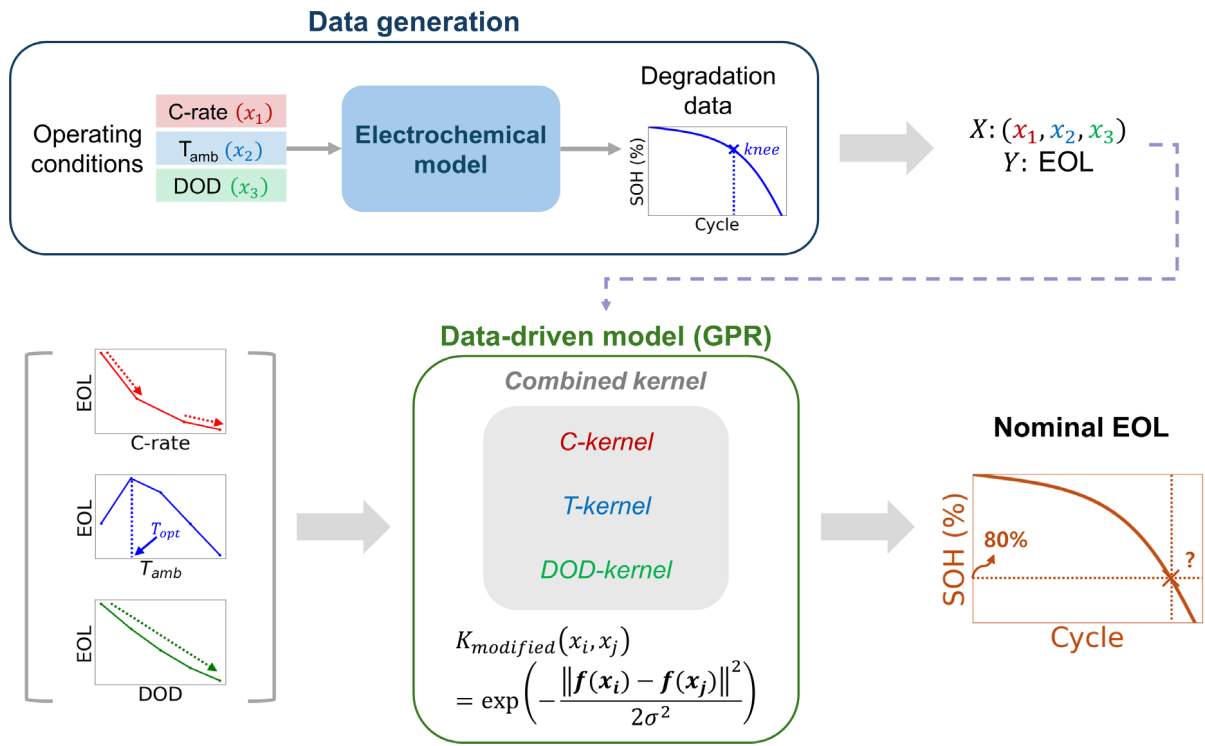

Fig. 1. Model architecture used in this work.

## 2. Data Generation

Let us first describe the data used in this study. Although many operating conditions affect the cycle life of a cell, this study focuses on the C-rate of the CC charging process, ambient temperature ($T_{amb}$), and cyclic DOD as the target operating conditions. These three operating conditions are known to impact degradation significantly and can be easily adjusted by users or according to different situations in real-world battery applications [23-26]. While the cycling data obtained from different operating conditions are required, such data are not readily available in the literature. Thus, this study utilizes synthetic data generated from a pseudo two-dimensional model, following the procedure given in Torchio et al. [27]. In this model, SEI formation and lithium plating are the degradation mechanisms causing capacity fade. Degradation in the early cycles is mainly caused by SEI formation, and the abrupt nonlinear fade that occurs afterward is mainly due to lithium plating [7]. A detailed discussion of both aging models can be found in [7, 28].

The cell to be simulated in this study is a $LiCoO_2$/ graphite pouch cell, and the model parameters are taken from the references as summarized in the appendix [7, 27, 28]. In all the cycles of one experiment, the operating conditions are fixed, and a cycle takes the form of *Discharge-Rest-CC charge-CV charge-Rest*. Discharging is performed until the cutoff SOC is reached with a CC discharging of 1C, which is equivalent to the current density of 29.23A/m². CC charging is performed until the cell voltage reaches 4.1V, which is equivalent to 80% SOC at the open circuit voltage of a fresh cell. The charging rate of the CC process is one of the target operating conditions. CV charging is performed at 4.1V until the cell reaches the upper limit SOC or the applied current is 0.1C or less. The two rest periods take 2000 seconds each to allow sufficient diffusion of lithium ions inside the cell. This cycle repeats until the cell reaches EOL, which in this study is defined as the point at which 80% SOH is reached. The operation of one cycle at 1C, 25°C and 80% DOD can be seen in Fig. 2 as a plot of V and I over time.

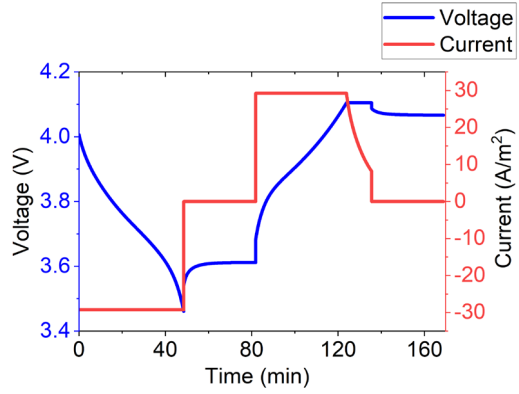

Fig. 2. Voltage and current over time in a cycle at 1C, 25℃ and 80% DOD.

A total of 100 experiments are conducted using four values of C-rate, five values of $T_{amb}$, and five values of DOD as summarized in Table 1. C-rate ranges from 1C, which is a relatively mild charging rate, to 2C, which can be considered fast charging. $T_{amb}$ is based on room temperature of 25℃, with the same range of low (5℃) and high (45℃). For cyclic DOD, to minimize the effect of SOC, medium SOC is fixed at 50%. For example, an experimental condition of 40% DOD means cycling from the 30% SOC to 70% SOC.

Table 1. Simulated OC values with EM.

| OC | Unit | Values |
| --- | --- | --- |
| C-rate (CC) | [C] | 1.0, 1.3, 1.7, 2.0 |
| $T_{amb}$ | [℃] | 5, 15, 25, 35, 45 |
| DOD | [%] | 40, 50, 60, 70, 80 |

# 3.  Methods

## 3.1.  Gaussian Process Regression

GPR is a probabilistic machine learning method based on Bayes' theorem [29]. It assumes that all data follow a Gaussian process with no prior information and converts this prior distribution to a posterior distribution, which is the conditional distribution under the observations. The prior distribution of a function $f(x)$ when it follows a Gaussian process is as follows:

$$f(x) \sim GP(m(x), k(x, x')) \tag{1}$$

where $m(x)$ is the mean function and $k(x, x')$ is the covariance function, also known as the kernel function. This kernel function reflects the similarities between the observations and between the observations and the predicted values. In GPR, the mean function of the observations is usually assumed to be zero for simplicity. The covariance term is mainly composed of the covariance function and noise term, denoted by $\sigma_n^2 \mathbf{I}_n$:

$$\mathbf{y} \sim N(0, \mathbf{K}(\mathbf{x}, \mathbf{x}) + \sigma_n^2 \mathbf{I}_n) \tag{2}$$

where $\mathbf{y}$ is the output.

For the prediction of the output $\mathbf{y}^*$ for a new dataset $\mathbf{x}^*$, the joint distribution with the prior distribution is:

$$\begin{bmatrix} \mathbf{y} \\ \mathbf{y}^* \end{bmatrix} \sim \mathcal{N}\left(0, \begin{bmatrix} \mathbf{K}(\mathbf{x}, \mathbf{x}) + \sigma_n^2 \mathbf{I}_n & \mathbf{K}(\mathbf{x}, \mathbf{x}^*) \\ \mathbf{K}(\mathbf{x}^*, \mathbf{x})^T & \mathbf{K}(\mathbf{x}^*, \mathbf{x}^*) \end{bmatrix} \right) \tag{3}$$

Based on this joint distribution, the posterior distribution of the new dataset $\mathbf{y}^*$ can be represented as:

$$p(\mathbf{y}^* | \mathbf{x}, \mathbf{y}, \mathbf{x}^*) = \mathcal{N}(\mathbf{y}^* | \bar{\mathbf{y}}^*, \text{cov}(\mathbf{y}^*)) \tag{4}$$

where

$$\bar{\mathbf{y}}^* = \mathbf{K}(\mathbf{x}, \mathbf{x}^*)^T [\mathbf{K}(\mathbf{x}, \mathbf{x}) + \sigma_n^2 \mathbf{I}_n]^{-1} \mathbf{y} \tag{5}$$

$$\text{cov}(\mathbf{y}^*) = \mathbf{K}(\mathbf{x}^*, \mathbf{x}^*) - \mathbf{K}(\mathbf{x}, \mathbf{x}^*)^T [\mathbf{K}(\mathbf{x}, \mathbf{x}) + \sigma_n^2 \mathbf{I}_n]^{-1} \mathbf{K}(\mathbf{x}, \mathbf{x}^*) \tag{6}$$

$\bar{\mathbf{y}}^*$ is the mean value of the predicted $\mathbf{y}^*$ on the test set $\mathbf{x}^*$ and $\text{cov}(\mathbf{y}^*)$ is the variance matrix of the predicted value. This matrix can be used to quantify the uncertainty of the predicted value by finding the confidence interval of the predicted value.

The performance of GPR highly depends on the selection of kernels, which capture the similarity between data. The most commonly used kernel is the radial basis function (RBF), which quantifies the similarity between datasets by the following equation [29]:

$$k(x_i, x_j) = \sigma^2 \exp\left(-\frac{d(x_i, x_j)^2}{2l^2}\right) \tag{7}$$

where $\sigma$ represents the amplitude of covariance, $d$ is the Euclidean distance, and $l$ is a parameter that indicates the length scale of the kernel, which can be used as a scalar in an isotropic kernel or as a vector in an anisotropic kernel that applies different length scales depending on input features. Although the RBF kernel is widely used, it typically performs poorly on nonlinear data, and prediction performance can be particularly poor when the data are of high dimensions. Therefore, to achieve high prediction performance on nonlinear data, the kernel must be properly tuned considering the characteristics of the data.

### 3.2. GPR Kernel Development

#### 3.2.1. C Kernel

Since EOL decreases with increasing C-rate monotonically and exponentially, the following characteristics are desired for the C kernel:

- The closer the distance between data, the higher the kernel value.

- Higher C-rate data has a higher kernel value when separated by the same distance.

The second characteristic is crucial to capture the exponential nature of the data. For example, when trying to predict the EOL value at 2C, a conventional RBF kernel would give equal weight to the data at 1C and 3C, which are at the same distance from 2C. However, the EOL at 2C is actually much closer to that at 3C than at 1C. To account for this nonlinearity, the distance of the reciprocal C-rate is used rather than the distance of the C-rate itself so that higher C-rates have larger kernel values. The C-rate kernel that reflects these characteristics is defined as:

$$k_C(x_i, x_j) = \exp\left(-\frac{\left\|\frac{1}{x_i} - \frac{1}{x_j}\right\|^2}{2\sigma_C^2}\right) \quad (8)$$

where $\sigma_C$ is the hyperparameter representing the length scale.

To examine the capability of the proposed C kernel, a pre-test is conducted with the C-rate vs EOL data provided in Wang et al. [26]. The empirical expression for capacity loss in each cycle obtained from the study was used to obtain the EOL as a function of C-rate in the range between 1C to 3C. The empirical equation used has the form of $Q_{loss} = B \cdot \exp(A \cdot \text{C-rate}) \cdot (\text{cycle\_number})^D$, where $A, B$ and $D$ are fitting parameters taking values of 0.7, 0.3, and 0.6, respectively.

This data is utilized in the GPR models with the modified C kernel and RBF kernel, and the two EOL prediction results are compared. As shown in Fig. 3(a), the conventional RBF predicts the EOL at 1.5C as a linear combination between the EOL values at 1C and 2C and performs relatively poorly. On the other hand, the exponential relationship between C-rate and EOL is well found by the proposed C kernel, and the prediction error is smaller.

This prediction performance can be further analyzed by comparing the actual correlation of EOL in each operating condition range to the kernel values between operating conditions as depicted in Fig. 4. The gray solid line is the actual EOL correlation value at a given C-rate, with values closer to 1 for similar EOL values. The red and green dashed lines represent the kernel values, and the closer the kernel value is to the EOL correlation value, the better the prediction performance. Unlike the RBF kernel, which only utilizes the physical distance between C-rate values, the proposed kernel better reflects the exponential relationship between C-rate and EOL. The average error between the actual EOL correlation value and the kernel value for each case can be found in Table 2, which shows that the proposed kernel reduces the error significantly.

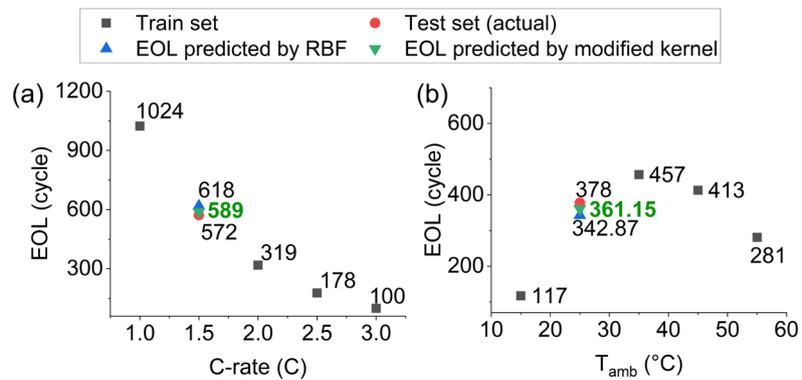

Fig. 3. Comparison of EOL prediction pre-test results when using RBF kernel and modified kernel. (a) C kernel, (b) T kernel.

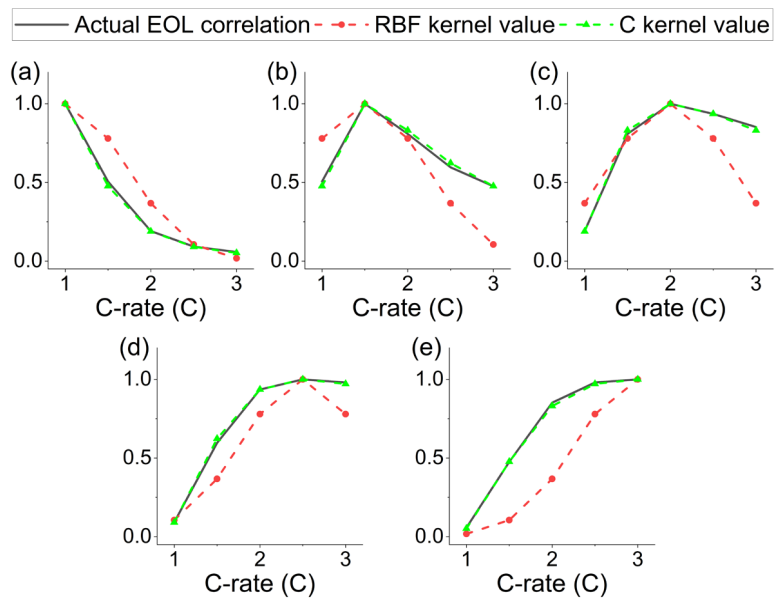

Fig. 4. Comparison of actual EOL correlation values and kernel values of the pre-test for all C-rate. (a) 1C, (b) 1.5C, (c) 2C, (d) 2.5C, (e) 3C.

### 3.2.2. T Kernel

For $T_{amb}$, due to the existence of an optimal temperature, the following characteristics need to be

considered:

- The closer the distance to the optimal temperature, the higher the kernel value.
- If they are the same distance away from the optimal temperature, the higher the temperature, the higher the kernel value.

If the optimal temperature is 25°C and the EOL at 20°C is predicted, an RBF kernel that considers the Euclidean distance will give the same weights to the data at 10°C and 30°C. However, since the optimum point exists at 25°C, it can be expected the actual EOL at 20°C to be much more similar to 30°C than 10°C. Weighting by comparing deviation from the optimal temperature, rather than simply measuring distance, can better reflect this characteristic. Additionally, the slope of the Arrhenius plot is much steeper at low temperatures compared to high temperatures, even when the deviations from the optimal temperature are the same. Therefore, higher temperatures should be weighted more heavily to account for this difference. Considering these characteristics, the following T kernel is proposed:

$$k_T(x_i, x_j) = \exp\left(-\frac{\left\|\frac{\|x_i - T_{opt}\|}{x_i + C_T} - \frac{\|x_j - T_{opt}\|}{x_j + C_T}\right\|^2}{2\sigma_C^2}\right) \tag{9}$$

where $T_{opt}$ is the optimal temperature, and $\sigma_T$ and $C_T$ are hyperparameters representing the length scale and the correction value, respectively. $T_{opt}$ is the temperature that maximizes the EOL at a specific C-rate and DOD, which changes with the C-rate and DOD. The difference between $T_{amb}$ and $T_{opt}$ is divided by $x + C_T$ to compensate for the greater weighting of higher temperatures. To avoid problems with negative T values, all values used by the proposed T kernel have the unit of K.

The performance of the proposed T kernel is compared with the RBF kernel by a pre-test using T vs. EOL data at 0.4C from Fig. 7(d) of Kucinskis et al. [25]. Since there are only data at 15, 25, 35, 45, and 55°C, it is assumed that 35°C is the optimal temperature with the maximum EOL among five conditions. As shown in Fig. 3(b), the traditional RBF kernel, which only depends on the distance between temperatures, predicts EOL that is very different from the actual value. The proposed T kernel performs much better, as it is able to reflect that the longest EOL is found at 35℃.

Fig. 5 shows the actual EOL correlation for each $T_{amb}$ versus the kernel value for all $T_{amb}$. Similar to the C kernel, the values of the proposed T kernel are much closer to the EOL correlation values than the RBF kernel across all $T_{amb}$ ranges. In particular, in Fig. 5(b), where the value at 25℃ is compared to the value at all other ranges, the RBF kernel has very different kernel values in the range above 35℃, which is the optimal temperature. However, the proposed T kernel has higher kernel values at 25 and 45℃ with the same distance from 35℃ and has a generally higher kernel value at high temperatures. This suggests that the T kernel is expected to have better prediction performance than the RBF kernel.

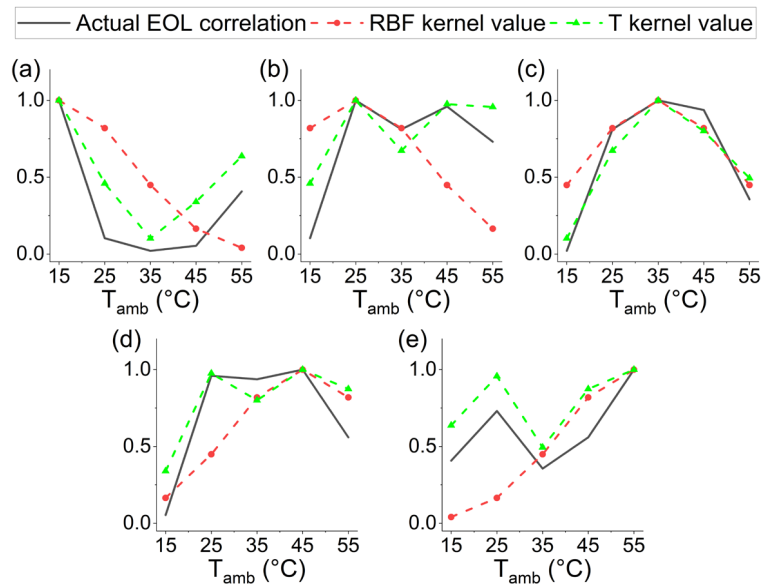

Fig. 5. Comparison of actual EOL correlation values and kernel values of the pre-test for all $T_{amb}$. (a) 15℃, (b) 25℃, (c) 35℃, (d) 45℃, (e) 55℃.

Table 2 compares the error between kernel value and EOL correlation by the kernel, illustrating the advantages of the proposed C and T kernels. However, unlike the C kernel, where the error converges to almost zero, the T kernel still has a rather large error. This is presumably caused by the value used for $T_{opt}$, which was determined by a limited amount of data. Still, in reality, $T_{opt}$ is very likely to be in unexplored conditions. This illustrates the importance of properly exploring $T_{opt}$, which will be discussed in more detail in Chapter 4.

Table 2. Error between actual EOL correlation and kernel values of the pre-test depending on the kernel and the improvement over the proposed kernel.

| C-rate (C) | RBF | C kernel | Improvement rate (%) |
|---|---|---|---|
| 1 | 0.1006 | 0.0078 | **92.26** |
| 1.5 | 0.1801 | 0.0159 | **91.14** |
| 2 | 0.1695 | 0.0093 | **94.50** |
| 2.5 | 0.1199 | 0.0074 | **93.80** |
| 3 | 0.2191 | 0.0075 | **96.56** |
| $T_{amb}$ (°C) | RBF | C kernel | Improvement rate (%) |
| 15 | 0.3244 | 0.1902 | **41.38** |
| 25 | 0.3596 | 0.1475 | **58.99** |
| 35 | 0.1293 | 0.0991 | **23.37** |
| 45 | 0.2000 | 0.1506 | **24.72** |
| 55 | 0.2570 | 0.1814 | **29.44** |

### 3.2.3. Combined Kernel

Similar to the C-rate, DOD has an exponential relationship with EOL. However, in the 40-80% DOD range used in this study, EOL is expected to decrease almost linearly with increasing DOD. Thus, the conventional RBF kernel is applied for DOD, which takes the following form:

$$k_{DOD}(x_i, x_j) = \exp\left(-\frac{\|x_i - x_j\|^2}{2\sigma_{DOD}^2}\right) \quad (10)$$

By combining the kernels proposed above, the effects of operating conditions can be simultaneously considered in an environment where all three conditions change together. The final kernel is as below:

$$k(\mathbf{x}_i, \mathbf{x}_j) = C \exp\left(-\frac{\left\|\frac{1}{x_{1,i}} - \frac{1}{x_{1,j}}\right\|^2}{2\sigma_C^2}\right)$$

$$\times \exp\left(-\frac{\left\|\frac{\|x_{2,i} - T_{opt}\|}{x_{2,i} + C_T} - \frac{\|x_{2,j} - T_{opt}\|}{x_{2,j} + C_T}\right\|^2}{2\sigma_C^2}\right) \quad (11)$$

$$\times \exp\left(-\frac{\|x_i - x_j\|^2}{2\sigma_{DOD}^2}\right) + \sigma_n \mathbf{I}$$

where $C$ is a hyperparameter expressing the amplitude of the kernel, and the last term reflects the noise. The $\mathbf{x}$ entering the kernel is a vector of $\mathbf{x} = (x_1, x_2, x_3)$ where $x_1$ is C-rate, $x_2$ is $T_{amb}$, and $x_3$ is DOD.

## 4. Results and Discussion

In this section, the performance of the final kernel is evaluated in comparison with the conventional RBF kernel in the main task, predicting the nominal EOL of LIBs generated by the P2D model as OC values in Table 1. The reliability of the predictions is also discussed by quantifying the uncertainty of the predicted EOL. For the evaluation, root mean square error (RMSE) and mean absolute percentage error (MAPE) are used:

$$\text{RMSE} = \sqrt{\frac{1}{N}\sum_{i=1}^{N}(y_i - \hat{y}_i)^2} \tag{12}$$

$$\text{MAPE} = \frac{1}{N}\sum_{i=1}^{N}\frac{|y_i - \hat{y}_i|}{y_i} \tag{13}$$

where $y_i$ is the actual value, $\hat{y}_i$ is the predicted value, and $N$ is the number of datasets.

### 4.1. Effect of the C Kernel

The prediction performance is compared as a function of C-rate to see if the C kernel captures the tendency of EOL correctly. The average prediction error at each C-rate is compared across the 20 test sets. To avoid biased results caused by the random split, the results of 5 random states are averaged.

From Fig. 6, it can be seen that the proposed C kernel outperforms the RBF kernel in all the C-rate ranges. For RMSE, the traditional RBF kernel has an error of 56.79, 35.93, 33.02, and 37.86 cycles for 1C, 1.3C, 1.7C, and 2C, respectively. The C kernel significantly reduces these errors to 53.16, 19.38, 13.26, and 26.97 cycles, respectively. The RMSE at a low C-rate is large because the EOL value is larger for lower C-rates. In order to analyze the effect more accurately, MAPE is also calculated. The MAPE values are 2.51, 2.65, 2.87, and 3.84% at 1, 1.3, 1.7, and 2C, respectively, using the RBF kernel. It can be seen that the error increases with increasing C-rate because the RBF kernel only relies on the distance between C-rate values. However, the proposed C kernel weights the high C-rate more heavily to reduce the MAPE to 2.46, 1.50, 1.30, and 3.12%, respectively.

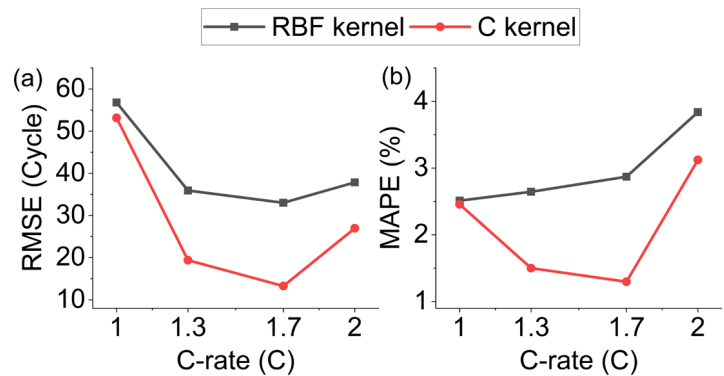

Fig. 6. EOL prediction error by kernel type across all C-rate ranges of test sets. (a) RMSE, (b) MAPE.

## 4.2. Effect of the T kernel

It is well known that $T_{opt}$ is affected by C-rate and DOD, and thus, it needs to be represented as a function of these operating conditions. However, C-rate and DOD influence $T_{opt}$ in nonlinear manners, and exact empirical expressions are unknown. Thus, in this study, polynomials of different degrees are tested to identify the simplest expression with high accuracy. Since there are many parameters in the first and second-order equations, $T_{opt}$ is estimated from training data to determine ranges for the parameters, and then grid search is used to find the final parameters. The EOL prediction error using the optimal expression obtained by progressively shrinking the grid for effective search is shown in Fig. 7(a).

The lowest RMSE of 80.34 cycles is found when $T_{opt}$ is 18.9°C, assuming zero-order, i.e., $T_{opt}$ is constant, which is relatively large. The first-order polynomial has the lowest error of 52.76 cycles when $T_{opt} = -9.04 + (15.90 \cdot \text{C-rate}) + (0.20 \cdot \text{DOD})$, which is still higher than the error of the case using conventional RBF kernel. Therefore, the second-order equation with an RMSE of 30.54 cycles is used, which is:

$$T_{opt} = -48.15 + (48.4 \cdot \text{C-rate}) + (0.77 \cdot \text{DOD}) + (-9.52 \cdot \text{C-rate}^2) + (-0.07 \cdot \text{C-rate} \cdot \text{DOD}) + (-3.93 \cdot 10^{-3} \cdot \text{DOD}^2) \tag{14}$$

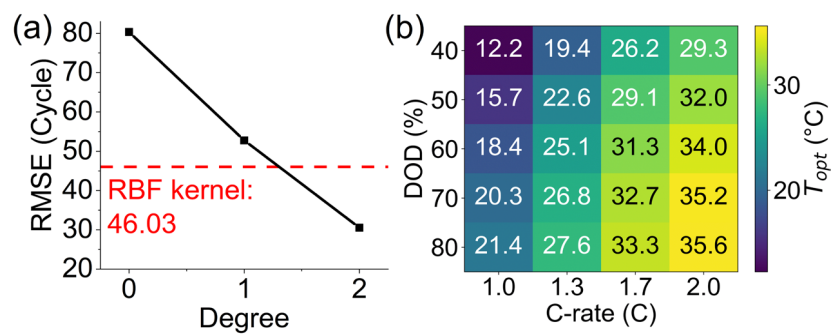

Fig. 7. (a) RMSE of EOL prediction based on the degree of the $T_{opt}$ expression. (Red dashed line: RMSE obtained using the RBF kernel.), (b) Calculated $T_{opt}$ as a function of C-rate and DOD.

Fig. 7(b) shows the results of predicting EOL with the proposed T kernel based on the $T_{opt}$

expression above. In all the T ranges, the proposed T kernel has higher prediction accuracy than the RBF kernel. The RMSE values for the RBF kernel are 71.50, 39.49, 20.81, 17.42 and 32.08 cycles for 5, 15, 25, 35, and 45°C, respectively, with large errors at the two endpoints. This is because the RBF kernel ignores the existence of an optimal temperature and only relies on the distance between temperatures. On the other hand, the proposed T kernel reduces the error significantly to 47.07, 23.73, 14.99, 16.27 and 16.51 cycles, respectively, and the prediction performances at 5 and 45°C are especially improved. Using the RBF, the MAPE values are 5.83, 2.99, 1.65, 1.60 and 2.10%, with large errors at 5°C. On the other hand, it can be seen that the proposed kernel reduces the errors to 4.62, 1.85, 1.39, 1.49 and 1.25%, respectively. A smaller improvement was obtained at 5°C due to the effect of lithium plating at low temperatures, which causes the EOL to decrease rapidly.

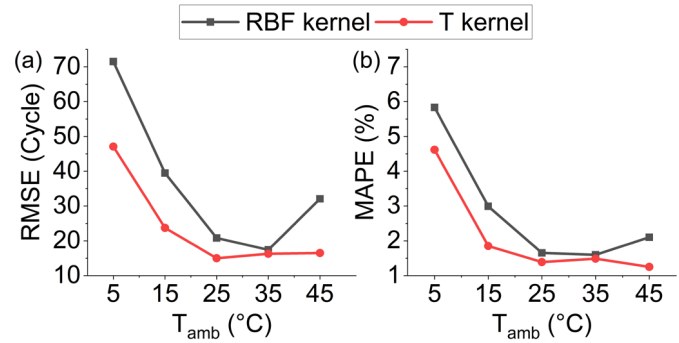

Fig. 8. EOL prediction error by kernel type across all $T_{amb}$ ranges. (a) RMSE, (b) MAPE.

### 4.3. Effect of the Combined Kernel

In this section, the prediction performance of using the traditional RBF kernel is compared with the case using the C kernel alone, the T kernel alone, and the combined C and T kernels. Hyperparameters of the original and modified kernels are determined by the values in Table 3.

Table 3. Optimal parameters of kernels.

| Kernel type | C | $\sigma_C$ | $\sigma_T$ | $\sigma_{DOD}$ | Noise level |
|---|---|---|---|---|---|
| Conventional RBF | $31.6^2$ | 0.313 | 19.82 | 18.40 | 1 |
| Modified kernel | $31.6^2$ | 0.223 | 0.255 | 15.70 | 1 |

Fig. 9 shows the prediction results for each kernel, comparing the predicted EOL to the actual EOL. The solid line shows the case of perfect prediction. When using the traditional RBF kernel, the error is

64.24 cycles for the entire test set, and the prediction performance is relatively poor. Using the C kernel alone or the T kernel alone, the performance is much better than the RBF kernel, with RMSEs of 43.87 and 24.6 cycles, respectively. In Fig. 9(d), the combined kernel is used, and the highest prediction performance with the RMSE of 20.42 cycles is achieved. Table 4 shows the overall prediction error on the test set of 5 randomly split datasets. In all the cases, the proposed kernels are more effective than the RBF kernel, and the combined kernel is more effective than the C kernel or T kernel alone.

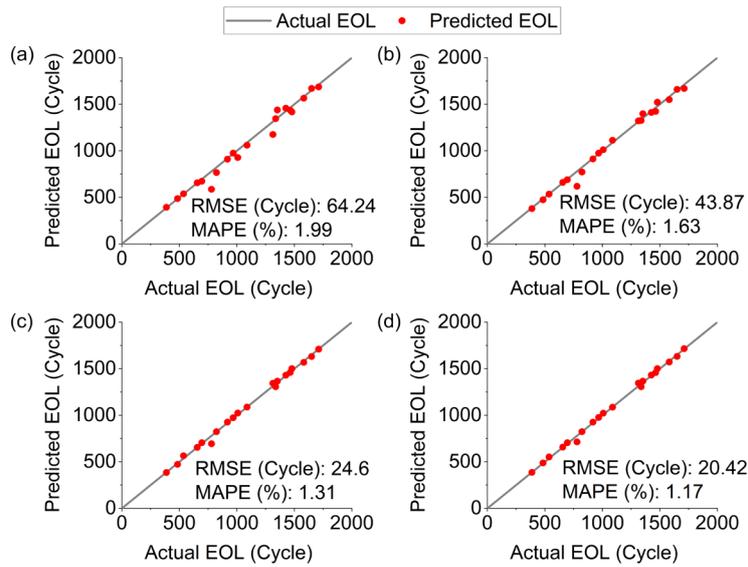

Fig. 9. Comparison between actual EOL and predicted EOL according to the kernel. (a) RBF kernel, (b) C kernel, (c) T kernel, (d) Combined kernel.

Table 4. EOL prediction error and uncertainty of the prediction depending on the kernel.

|  | RBF | C kernel | T kernel | **Combined** |
|---|---|---|---|---|
| **Prediction error** | | | | |
| RMSE (cycle) | 46.03 | 34.44 | 30.54 | **24.57** |
| MAPE (%) | 1.75 | 1.42 | 1.52 | **1.28** |
| **Uncertainty** | | | | |
| Covariance (cycle) | 4.50 | 3.58 | 3.32 | **2.60** |

Since the GPR model obtains EOL prediction results as a distribution rather than a single value, it is possible to quantify the uncertainty of the predicted value, which can be seen in Table 4. Conventional RBF, C kernel, T kernel, and combined kernel have progressively smaller uncertainties. Small uncertainty means that the GPR model captures the trend of EOL variation with operating conditions well. This leads to the conclusion that the proposed kernels are much more efficient and reliable for

EOL prediction compared to the RBF kernel.

## 4.4. Further Discussions

Although this study showed high accuracy in nominal EOL prediction by appropriately modifying the kernel of the GPR model, some limitations exist. First, the intrinsic cell-to-cell variability was not considered, which is crucial to determining the exact value of EOL. Also, there exist other operating conditions not considered in this study, which can affect the EOL. It is our goal to develop a more generalized EOL prediction methodology by considering more operating conditions and adopting residual learning to address cell-to-cell variability.

In addition, due to the lack of prior information about $T_{opt}$, the expression is restricted to a polynomial form, which may have hindered the prediction performance. Also, to find the parameters of the expression, we performed a grid search using the data from the train set as initial values, which may not be optimal. Once more information becomes available on $T_{opt}$, the accuracy of $T_{opt}$ can be improved, making the model more useful and reliable.

## 5. Conclusion

Accurate EOL prediction is important for LIBs because they become less safe as they approach their EOL. Previous studies have not separately considered the effects of operating conditions and intrinsic cell characteristics, leading to expensive battery cycle life predictions. Starting from the fact that operating conditions have a stronger impact on degradation, this study proposes a GPR model to predict the nominal EOL of LIB very efficiently and accurately using only three operating conditions: C-rate, ambient temperature, and DOD. To overcome the problem of battery data scarcity, a P2D model was combined with SEI formation and lithium plating models to produce 100 nonlinear degradation datasets under various operating conditions. A novel kernel is proposed to reflect the relationship between C-rate and EOL, and between ambient temperature and EOL, which cannot be captured by the RBF kernel effectively. The proposed kernel does not require cycling experiments because it uses only the operating conditions as the inputs. By comparing the similarity of the kernel value with the correlation of the actual EOL, it was found that the proposed kernel is much better at inferring the similarity between data. As a result, the RMSE of the EOL prediction of the test set was reduced by 25.18% using the C kernel alone, 33.65% using the T kernel alone, and 46.62% using the combined kernel compared to the RBF kernel. The uncertainty of the prediction result was also reduced by using the proposed kernel.

# References

[1] Hu X, Cao D, Egardt B. Condition monitoring in advanced battery management systems: Moving horizon estimation using a reduced electrochemical model. IEEE/ASME Transactions on Mechatronics. 2017;23:167-78.

[2] Blomgren GE. The development and future of lithium ion batteries. Journal of The Electrochemical Society. 2016;164:A5019.

[3] Barré A, Deguilhem B, Grolleau S, Gérard M, Suard F, Riu D. A review on lithium-ion battery ageing mechanisms and estimations for automotive applications. Journal of power sources. 2013;241:680-9.

[4] Ng KS, Moo C-S, Chen Y-P, Hsieh Y-C. Enhanced coulomb counting method for estimating state-of-charge and state-of-health of lithium-ion batteries. Applied energy. 2009;86:1506-11.

[5] Edge JS, O'Kane S, Prosser R, Kirkaldy ND, Patel AN, Hales A, et al. Lithium ion battery degradation: what you need to know. Physical Chemistry Chemical Physics. 2021;23:8200-21.

[6] Devie A, Dubarry M. Durability and reliability of electric vehicle batteries under electric utility grid operations. part 1: Cell-to-cell variations and preliminary testing. Batteries. 2016;2:28.

[7] Yang X-G, Leng Y, Zhang G, Ge S, Wang C-Y. Modeling of lithium plating induced aging of lithium-ion batteries: Transition from linear to nonlinear aging. Journal of Power Sources. 2017;360:28-40.

[8] Waldmann T, Wilka M, Kasper M, Fleischhammer M, Wohlfahrt-Mehrens M. Temperature dependent ageing mechanisms in Lithium-ion batteries–A Post-Mortem study. Journal of power sources. 2014;262:129-35.

[9] Pang X, Zhong S, Wang Y, Yang W, Zheng W, Sun G. A Review on the Prediction of Health State and Serving Life of Lithium-Ion Batteries. The Chemical Record. 2022;22:e202200131.

[10] Che Y, Hu X, Lin X, Guo J, Teodorescu R. Health prognostics for lithium-ion batteries: mechanisms, methods, and prospects. Energy & Environmental Science. 2023;16:338-71.

[11] Li Y, Liu K, Foley AM, Zülke A, Berecibar M, Nanini-Maury E, et al. Data-driven health estimation and lifetime prediction of lithium-ion batteries: A review. Renewable and sustainable energy reviews. 2019;113:109254.

[12] Zhang Y, Li Y-F. Prognostics and health management of Lithium-ion battery using deep learning methods: A review. Renewable and sustainable energy reviews. 2022;161:112282.

[13] Aitio A, Howey DA. Predicting battery end of life from solar off-grid system field data using machine learning. Joule. 2021;5:3204-20.

[14] Hsu C-W, Xiong R, Chen N-Y, Li J, Tsou N-T. Deep neural network battery life and voltage prediction by using data of one cycle only. Applied Energy. 2022;306:118134.

[15] Severson KA, Attia PM, Jin N, Perkins N, Jiang B, Yang Z, et al. Data-driven prediction of battery cycle life before capacity degradation. Nature Energy. 2019;4:383-91.

[16] Ma G, Zhang Y, Cheng C, Zhou B, Hu P, Yuan Y. Remaining useful life prediction of lithium-ion batteries based on false nearest neighbors and a hybrid neural network. Applied Energy. 2019;253:113626.

[17] Lee J, Lee JH. Simultaneous extraction of intra-and inter-cycle features for predicting lithium-ion battery's knees using convolutional and recurrent neural networks. Applied Energy. 2024;356:122399.

[18] Richardson RR, Birkl CR, Osborne MA, Howey DA. Gaussian process regression for in situ capacity estimation of lithium-ion batteries. IEEE Transactions on Industrial Informatics. 2018;15:127-38.

[19] Lucu M, Martinez-Laserna E, Gandiaga I, Camblong H. A critical review on self-adaptive Li-ion battery ageing models. Journal of Power Sources. 2018;401:85-101.

[20] Yang D, Zhang X, Pan R, Wang Y, Chen Z. A novel Gaussian process regression model for state-of-health estimation of lithium-ion battery using charging curve. Journal of Power Sources. 2018;384:387-95.

[21] Liu K, Li Y, Hu X, Lucu M, Widanage WD. Gaussian process regression with automatic relevance determination kernel for calendar aging prediction of lithium-ion batteries. IEEE Transactions

on Industrial Informatics. 2019;16:3767-77.

[22] Liu K, Hu X, Wei Z, Li Y, Jiang Y. Modified Gaussian process regression models for cyclic capacity prediction of lithium-ion batteries. IEEE Transactions on Transportation Electrification. 2019;5:1225-36.

[23] Lin X, Khosravinia K, Hu X, Li J, Lu W. Lithium plating mechanism, detection, and mitigation in lithium-ion batteries. Progress in Energy and Combustion Science. 2021;87:100953.

[24] Choi SS, Lim HS. Factors that affect cycle-life and possible degradation mechanisms of a Li-ion cell based on LiCoO2. Journal of Power Sources. 2002;111:130-6.

[25] Kucinskis G, Bozorgchenani M, Feinauer M, Kasper M, Wohlfahrt-Mehrens M, Waldmann T. Arrhenius plots for Li-ion battery ageing as a function of temperature, C-rate, and ageing state–An experimental study. Journal of Power Sources. 2022;549:232129.

[26] Wang J, Liu P, Hicks-Garner J, Sherman E, Soukiazian S, Verbrugge M, et al. Cycle-life model for graphite-LiFePO4 cells. Journal of power sources. 2011;196:3942-8.

[27] Torchio M, Magni L, Gopaluni RB, Braatz RD, Raimondo DM. Lionsimba: a matlab framework based on a finite volume model suitable for li-ion battery design, simulation, and control. Journal of The Electrochemical Society. 2016;163:A1192.

[28] Ramadass P, Haran B, Gomadam PM, White R, Popov BN. Development of first principles capacity fade model for Li-ion cells. Journal of the Electrochemical Society. 2004;151:A196.

[29] Rasmussen CE. Gaussian processes in machine learning. Summer school on machine learning: Springer; 2003. p. 63-71.